\documentclass[10pt]{article}
\usepackage{amsmath, a4wide, amssymb, mathrsfs, bbm, amsthm, chicago}
\usepackage[english]{babel}
\usepackage[mathcal]{eucal}
\usepackage[title]{appendix}

\newcommand{\real}[1]{\ensuremath{\mathbb{R}^{#1}}}

\title{{\bfseries Foundations of Quantum Gravity : The Role of Principles Grounded in Empirical Reality}}
\author{M. Holman \\ Department of Physics and Astronomy, \hspace{0.05cm} Utrecht University, \\ Princetonplein 5,  \hspace{0.05cm} 3584 CC  Utrecht, \hspace{0.05cm} The Netherlands \\ e-mail : {\ttfamily m.holman@phys.uu.nl}}

\begin{document}
\maketitle 
\begin{abstract}
\noindent When attempting to assess the strengths and weaknesses of various principles in their potential role of guiding the 
formulation of a theory of quantum gravity, it is crucial to distinguish between principles which are strongly supported by 
\emph{empirical data} - either directly or indirectly - and principles which instead (merely) rely heavily on theoretical 
arguments for their justification.
Principles in the latter category are not necessarily invalid, but their a priori 
foundational significance should be regarded with due caution.
These remarks are illustrated in terms of the current standard models of cosmology and particle physics, as
well as their respective underlying theories, i.e. essentially general relativity and quantum (field) theory.
For instance, it is clear that both standard models are severely constrained by symmetry principles~: an effective 
homogeneity and isotropy of the known universe on the largest scales in the case of cosmology and an underlying exact 
gauge symmetry of nuclear and electromagnetic interactions in the case of particle physics.
However, in sharp contrast to the cosmological situation, where the relevant symmetry structure is more or less established
directly on observational grounds, all known, nontrivial arguments for the ``gauge principle'' are purely theoretical (and 
far less conclusive than usually advocated).
Similar remarks apply to the larger theoretical structures represented by general relativity and quantum (field) theory,
where - actual or potential - empirical principles, such as the (Einstein) equivalence principle or EPR-type nonlocality,
should be clearly differentiated from theoretical ones, such as general covariance or renormalizability.
It is argued that if history is to be of any guidance, the best chance to obtain the key structural features of a putative
quantum gravity theory is by deducing them, in some form, from the appropriate empirical principles (analogous to the
manner in which, say, the idea that gravitation is a curved spacetime phenomenon is arguably implied by the equivalence principle).
Theoretical principles may still be useful however in formulating a concrete theory (analogous to the manner in which, say,
a suitable form of general covariance can still act as a sieve for separating theories of gravity from one another).\\
It is subsequently argued that the appropriate empirical principles for deducing the key structural features of 
quantum gravity should at least include (i) quantum nonlocality, (ii) irreducible indeterminacy (or, essentially 
equivalently, given (i), relativistic causality), (iii) the thermodynamic arrow of time, (iv) homogeneity and isotropy 
of the observable universe on the largest scales.
In each case, it is explained - when appropriate - how the principle in question could be implemented mathematically in a 
theory of quantum gravity, why it is considered to be of fundamental significance and also why contemporary accounts of it are insufficient.
For instance, the high degree of uniformity observed in the Cosmic Microwave Background is usually regarded as theoretically 
problematic because of the existence of particle horizons, whereas the currently popular attempts to resolve this 
situation in terms of inflationary models are, for a number of reasons, less than satisfactory. 
However, rather than trying to account for the required empirical features \emph{dynamically}, an arguably 
much more fruitful approach consists in attempting to account for these features directly, in the form of a lawlike initial 
condition within a theory of quantum gravity.
\end{abstract}\enlargethispage*{5cm} 

\noindent {\bfseries Keywords} : quantum gravity, thermodynamic arrow of time, CMB uniformity,\\
\noindent quantum nonlocality, irreducible randomness.

\newpage

\begin{quote}
It may be that a real synthesis of quantum and relativity theories requires not just technical developments but radical
conceptual renewal.\\
J.S. Bell \citeyear{Bell1}
\end{quote}

\section{Introduction}\label{intro}

\noindent The problem of how to unify quantum theory and general relativity into a consistent theoretical framework
has baffled physicists for more than eight decades, but has become particularly pressing in recent years.
Each of the two theories in question is a fundamental theory of physics in its own right - each theory is, in Einstein's terminology,
a \emph{theory of principle}; see below - and is corroborated by a very impressive body of experimental evidence.
But physical reality constitutes an undivided whole and it obviously makes no sense to have \emph{two} distinct fundamental
theories for a single reality.
That physicists have been able to get away with this state of affairs for so long is simply due to the fact that the
intersection of the domains of applicability of the two theories is in some sense small and at any rate is usually thought
to lie far beyond anything technologically feasible, at present or in the nearby future.
Indeed, according to the received view, because of gravity's intrinsic weakness as compared to the nuclear and electromagnetic
interactions, general relativistic considerations play no role in quantum theory until particle scattering energies reach 
Planckian dimensions and, conversely, because quantum theory is at first glance effectively restricted to the (sub-)atomic 
regime, it can seemingly play no role in any nontrivial general relativistic effects, as these typically involve distance 
scales many orders of magnitude beyond the atomic scale.
But, apart from the questionable nature of these arguments, the lack of a unifying theoretical framework for the entire
realm of natural phenomena is arguably a serious problem from the perspective of the foundations of physical theory\footnote{That is,
the lack of such a framework \emph{is} a serious problem assuming an appropriate form of scientific realism (although that is 
of course the household doctrine of theoretical physicists working on these matters - at least when explaining 
their activities to the lay public or when writing up research grants !). It is not necessary however to assume realism to
point out the inadequacies of existing theories (see section \ref{nwthrarguments} for further discussion).}.\\
As said, both quantum theory and general relativity are theories of principle. This means that they are characterized by 
defining principles which are supposed to be valid universally, and which should apply, without 
restriction, to \emph{all} physical phenomena\footnote{See \citeN{Stachel}.} (such theories are to be distinguished from \emph{constructive theories},
 which are theories which attempt to explain a limited group of natural phenomena by means of some model or set 
of equations, but always so within the context of a specific theory of principle\footnote{For instance, the
$\mbox{SU}(3) \times \mbox{SU}(2) \times \mbox{U}(1)$ standard model of particle physics and quantum field theory 
models more generally are constructive theories within a quantum theoretical context (see subsection \ref{QTandstmod}),
while the so-called $\Lambda \mbox{CDM}$ ``concordance model'' is a constructive theory of cosmology within a general relativistic 
context (see subsection \ref{GRandstmod}). It is sometimes argued that general relativity is (at least partly) a 
constructive theory in view of the fact that it can be regarded as a classical field theory, but that position will 
not be adopted here. General relativity describes the properties of space and time in relation to those of ponderable 
matter and fields and as such it forms the general setting for particular constructive theories of ponderable 
matter and fields.}).
But it is not difficult to see that the two sets of principles associated with these theories are not fully inter-compatible.
For instance, according to general relativity, spacetime is a dynamical entity, whereas the standard unitary dynamics 
of states and observables in quantum theory is conditional upon a decomposition of this entity into ``space'' and ``time'',
which are moreover taken as non-dynamical. Similarly, according to 
quantum theory it makes sense to talk of a point mass being in a quantum superposition centred about two different spatial 
locations, whereas according to general relativity this makes no sense, since such a superposition would (in principle) 
entail a corresponding superposition of two spacetime geometries and there is no natural way to mathematically implement
such a superposition (i.e., to identify the points of two non-identical spacetimes with each other)\footnote{Cf. \citeANP{Penrose10} \citeyear{Penrose10,Penrose9}. See also \shortciteN{Karolyhazy}, \citeN{Diosi}.}.
Logic thus dictates that at least one set of principles cannot be the complete story and it is a matter of simple
historical fact that the ``mainstream'', often implicit, viewpoint in theoretical physics with respect to this issue has 
always been that the principles of general relativity are somehow to be subjugated to those of quantum theory
(in a sense, this view is of course also encapsulated in the very phrase ``quantization of gravity'')\footnote{It is somewhat of an irony
that as early as 1916, Einstein himself already argued for a modification of ``the new theory of gravitation'' 
\emph{because} of quantum theory (of which the general principles had yet to be formulated of course). See \citeN{Einstein2}.
For an explicit quantum-universality claim in this regard, see e.g. \citeN{Kiefer}.}.
As will be seen in the next two sections however, there are ample grounds to question the validity of this viewpoint (at the
very least, it is undeniable that it has so far failed to lead to a consistent resolution of the unification problem).
If the mainstream perspective is abandoned, the incompatibility of the two sets of principles raises some profound 
conceptual and methodological dilemmas. For instance, on the one hand, it might seem that the most important lesson that can be drawn from the historical 
development of general relativity is that the strong insistence on a principle of locality, in spite of a number of serious 
earlier obstacles, ultimately payed off\footnote{The notion of locality will be further explicated and differentiated in subsection \ref{qnonloc}
but the intuitive content of this notion should suffice for present purposes.}. 
On such an account, it would thus not seem unreasonable to expect that quantum theory will eventually also be seen to be
founded upon a more local, ``classical'' basis. On the other hand, it might be
asked on what precise grounds a physical theory is required to be local. One seemingly obvious answer is that \emph{if}
physical reality is structured that way, human intuition would presumably manage to latch onto such a structure somehow,  
and so the \emph{fact} that locality is a very intuitive notion would then (according to this line of reasoning) imply the validity
of some form of locality as a fundamental principle of nature. Yet, such a position appears dangerously close to the 
Kantian notion of the synthetic a priori, whereas the intuitive appeal of a locality principle seems equally well 
explainable on the basis that the structure of ``macroscopic'' physical reality happens to be local (to a large extent) 
and that human intuition is able to latch onto that structure through sense experience. But there would then be no 
guarantee that such intuition could be validly extended into the ``microscopic'' domain, not directly accessible to 
the senses. 
In fact, as recalled in more detail later, it is well established that an underlying local description in the case of quantum 
theory is \emph{impossible} for empirical reasons (at least if such a description is interpreted in terms of real spacetime 
phenomena).\\
On this basis, it is then perhaps tempting to conclude that the entire doctrine of locality - although extremely useful 
in arriving at the successful theories of electromagnetism and relativity - is ultimately untenable and that, because of this,
the principles of quantum theory carry more weight than those of general relativity. 
However, even if the locality of general relativity and classical electrodynamics ultimately turns out to be ``merely'' an emergent quality 
- as the present evidence strongly suggests - there is an important reason for attaching at least as much weight to the line of 
research that culminated in Einstein's general relativity and, consequently, also to the principles of general relativity themselves. 
This is the circumstance that the development of the theories of relativity - as well as that of classical electromagnetism - 
were strongly driven by the requirement that physical theories be \emph{intelligible}, whereas, by contrast, the development
of quantum theory was strongly dominated by an attitude which was essentially instrumentalist.\\ 
These remarks are further illustrated in subsections \ref{GRandstmod} and \ref{QTandstmod} respectively.
As to the unification problem, the upshot of the foregoing discussion is this.
It is clear that adoption of some set of guiding principles is necessary in order to have a realistic chance of 
formulating a concrete, consistent theory of \emph{quantum gravity} - i.e., an ``even-handed'' theory of principle
which contains both quantum theory and general relativity in the appropriate limits (and which, in particular, need thus
not fully comply with either theory).
But given the partial incompatibility of the limiting theories, as well as the plethora of different, often orthogonal
approaches that have already been attempted in this regard, which principles should be adopted ?
It is argued in section \ref{principles} that if history is to be of any guidance, the best chance to obtain the key 
structural features of a putative quantum gravity theory is by deducing them, in some form, from the appropriate ``empirical
principles'', i.e., principles which are strongly grounded in observational facts (as opposed to ``theoretical
principles'', which are not so grounded).
Four candidate empirical principles are subsequently identified in section \ref{QGprin}.
In each case, it is explained - when appropriate - how the principle in question could be implemented mathematically in a 
theory of quantum gravity, why it is deemed to be of fundamental significance and also why contemporary accounts of it are insufficient.
It is important to stress however that the principles listed are not necessarily foreseen to (all) play a role as \emph{postulates}
in a future theory of quantum gravity. Rather their roles may be restricted to function ``merely'' as \emph{guiding principles} (in the
same sense, for instance, as the equivalence principle crucially guided the formulation of general relativity, although it is not a
postulate of that theory).

\section{The Need for a New Theoretical Framework}\label{nwthrarguments}

\noindent As noted in the introduction, the standard approach to the quantum gravity problem assumes that general relativity
somehow emerges as the ``classical limit'' from a yet-to-be-obtained true \emph{quantum theory} of gravity of some sorts
(in the same manner, say, as classical electrodynamics is customarily thought to coincide with the ``classical limit'' of 
quantum electrodynamics).
Conceptually speaking, this standard approach fits well with (some version of) the doctrine of atomist reductionism
that underlies modern high-energy physics and according to which it is the supreme task of theoretical physics to
uncover the ultimate, i.e. smallest, constituents of matter and their governing principles\footnote{The general view 
towards the reductionism issue in this work is, first, that there are in fact several such issues, which are
unfortunately often grouped under the same banner however, and, second, that some notions of reductionism are reasonable,
or even necessary for science to be possible, while others are not.
For instance, there is a clear (and rather uncontroversial) sense in which a science such as physics lies at a lower 
level in the descriptional hierarchy than sciences such as chemistry or biology, but this does not mean that the latter 
can be fully reduced to physics - not even in principle. In fact, there are powerful reasons coming from within science 
to reject such a particular strong form of reductionism. Similar remarks incidentally apply to analogous hierarchical 
orderings \emph{within} physics. See \citeN{Holman2} for further discussion of these issues.}.
However, apart from the fact that there are a few well-known, general obstructions to atomist reductionism,
the standard approach is silent on how to interpret quantum theory, even though this is an issue central to its very 
viability in the first place~!
Indeed, according to Bohr's Copenhagen interpretation (or some modern version of it), which constituted the prevailing 
orthodoxy until relatively recently, there is only a quantum-mechanical ``description'' of things and no actual ``quantum world''.
Yet, it is actually very far from clear that such anti-realism can be made conceptually consistent with the type of
reductionism underlying high-energy physics (i.e., for what reason attempt to formulate a quantum-mechanical
\emph{description} of non-real structures supposedly relevant in some way at scales which are at least sixteen orders
of magnitude beyond currently accessible scales ? for what reason assume that such a description, granted that it can
be obtained, would in any meaningful sense be \emph{final} ?).
Moreover, even in the unlikely event that the philosophy underlying modern high-energy physics (as just portrayed) can
be made conceptually coherent with an anti-realist interpretation of quantum theory, there is the elementary observation
that, despite massive efforts for the past thirty years or so, a quantum theory of everything has remained elusive (at best),
so that it is natural to explore other possible directions.
On the other hand, on adopting a realist interpretation of quantum theory of some form, interpretational difficulties
also emerge. 
For instance, a question that immediately emerges when adopting a literal many worlds perspective is : why introduce
an extravagance of monstrous proportions, with no empirical ramifications whatsoever, just to account for a 
straightforward discrepancy between theory and observation ?
In a similar vein, any ``refined'' interpretation of quantum theory, in which the values of all observables always have 
determinate values (such as the de Broglie-Bohm pilot wave theory) necessarily conflicts with relativistic causality
and it becomes imperative to address the issue of whether initial conditions can always be ``reasonably'' chosen so as 
to avoid standard causal paradoxes (where people kill their own ancestors, for instance).
And of course, a perspective in which the state vector is objectively reduced naturally points to a theory beyond quantum
theory (disregarding far-fetched views in which for instance consciousness collapses the state vector).
That is, upon such a view it is natural to interpret state reduction and unitary dynamics
somehow as different aspects of the same fundamental dynamics.\\
This is not the place to go into the pros and cons of all the multifarious interpretations and sub-interpretations
of quantum theory (for some additional comments see subsections \ref{QTandstmod}, \ref{qnonloc} and \ref{qindet}), but the message should be 
clear. Unless it is indicated how to consistently interpret quantum theory, the mainstream view, according to which general 
relativity is to be somehow seen as the ``classical limit'' of a true microscopic theory, is at best good for heuristic 
purposes\footnote{In a sense, analogous remarks in fact apply to theories such as classical electrodynamics, given the
lack of well defined quantum versions of these theories at present.}.
Upon attempting to pursue a more even-handed approach to the quantum gravity problem, a question which immediately presents
itself is how to possibly uncover the postulates for a fundamental theory, when the principles of its two limiting theories
appear so disconnected from each other.
For instance, except for the postulated unitary dynamics and projective measurements, the axioms of quantum theory in their 
usual form make no reference to spacetime concepts, which makes it difficult to connect them to the foundations of relativity.
The resolution to this dichotomy lies of course in the recognition that within an even-handed approach, the principles 
of \emph{both} quantum theory and general relativity are likely to be subtly modified, re-appear in a very different form,
or even be replaced altogether (for instance, it will be argued later that there are good reasons to regard quantum
nonlocality of EPR type as a key principle of any reformulation of quantum theory relevant to the quantum gravity problem).
In fact, there are also motivations coming more or less from within each theory separately that point in this direction.
For instance, it is well known that general relativity predicts the existence of spacetime singularities under fairly 
generic and physically reasonable conditions\footnote{Cf. \citeN{HawPen}.} and that, in a sense, it thus contains its own 
seed of destruction. It is commonly assumed that any such singularities are removed upon including ``quantum gravity effects'' 
in an appropriate manner, but whether this happens through an ordinary quantization of classical structures such as loops
or strings (as occurs in the standard approaches), or through something more elaborate, it entails a breakdown of 
classical general relativistic structure.
Quantum theory, when considered as a theory of principle (i.e., as formulated in terms of its usual postulates involving an
abstract Hilbert space, linear operators on this space, etc.) is completely consistent, mathematically speaking (there
are a few subtleties - e.g. domain problems corresponding to unbounded self-adjoint operators - but these can be overcome
at the price of slight technical complication - e.g. through exponentiating self-adjoint operators to their unitary associates).
So, at first sight, it appears that this feature of quantum theory causes it to be slightly better off than general 
relativity. This way of representing things is somewhat deceptive however.
First, Einstein's equation certainly admits singularity-free solutions and general relativity, when considered as a
theory of principle, is certainly also fully consistent mathematically (as it should be, of course).
It is only when certain (arguably physically realistic) \emph{assumptions} are made on the stress-energy content of
spacetimes representing for instance black holes or the universe at large, that singularities are generically encountered.
Now, although it is also true that - \emph{unlike} the case of general relativity - a large number of physically relevant, 
mathematically consistent solutions to quantum theory is known to exist, these solutions are, with very 
few exceptions, all \emph{non-relativistic}.
They are therefore physically relevant only to the extent that relativistic effects can be safely ignored (something
which can usually be done in atomic physics, for instance, although even here it may be necessary to include
relativistic effects, such as fine structure, perturbatively, i.e., ``by hand'').
It is usually held that the incorporation of special relativistic structure into quantum theory requires the introduction
of quantum \emph{fields}\footnote{See in particular \citeN{Weinberg2}.
The point is that, although it is possible to write down relativistic wave equations (as famously first found by 
Schr\"odinger, Dirac and others and as later derived in a more systematic fashion by Wigner), such equations necessarily
pertain to \emph{fixed} numbers of particles and are thus unable to describe situations in which particles are annihilated
or created. Moreover, all these equations suffer from certain paradoxes associated with negative energy solutions (such as
Klein's paradox and Zitterbewegung) and more importantly, they give rise to discrepancies with experiments (as evidenced
by e.g. the Lamb shifts of atomic hydrogen).
These difficulties were essentially all resolved by re-interpreting the wavefunctions as classical fields, which were then to be 
subjected to a (second) quantization procedure.
Quantum field theory streamlines this process by essentially starting from a classical field theory and performing a
quantization procedure on this theory only once.}.
This common wisdom will not be called into question here, but it is noted that for a quantum theory of fields, the issue
of mathematical consistency changes rather drastically.
In fact, at present, no interacting, four-dimensional quantum field theories are known which are (a) mathematically consistent and (b)
demonstrably physically relevant (e.g. non-supersymmetric), while there is actually rather
strong theoretical evidence that any non-asymptotically free, four-dimensional interacting quantum field theory is mathematically \emph{in}consistent,
because of the expected presence of a Landau pole\footnote{That is, because of the probable presence of a \emph{finite}
energy scale for which the coupling constant of the theory diverges (\citeN{Landau}).
What can be stated with mathematical certainty is that perturbation theory unavoidably breaks down in the ultraviolet
for theories that are not asymptotically free (\citeN{GelLow}, \citeN{Weinberg3}).
Since all known renormalizability proofs are perturbative in character, this unfortunately complicates the issue of
determining the ultimate status of the Landau pole.\label{Landaupole}}.
So, any ``realistic'' quantum theory that nontrivially incorporates special relativity most presumably also carries its own seed of
destruction.
Again, it is commonly assumed that any inconsistencies are removed by including ``quantum gravity effects'' in an 
appropriate manner (for instance, through introduction of a short-distance cutoff associated with a fundamental
length scale), implying a breakdown of (special relativistic) quantum structure.
Of course, quantum theory is also under pressure because of the embarrassing lack of agreement among practising
physicists on what the theory is ultimately about and because of the earlier mentioned partial incompatibility of its
principles with those of general relativity. Concerning this latter point, the following observation is also useful.\\
In recent years some results have been obtained which seem to indicate that, while application of standard quantization
techniques to (the weak-field limit of) general relativity results in a theory that is not perturbatively renormalizable
(and hence - barring any deeper principle constraining the associated infinitely many free parameters to a finite
number - in a theory that fails to have any predictive power), there may nevertheless be reason to believe in the existence
of a stable ultraviolet fixed point of the renormalization group flow\footnote{This would mean that there \emph{is} 
some deeper principle and that effectively there are only finitely many free parameters. Following Weinberg, the theory 
is then said to be \emph{asymptotically safe}. For recent reviews of the question of whether quantum gravity meets this
criterion, see e.g. \citeN{NieReu} or \citeN{Percacci}.\label{asysafe}}.
However, what is not typically addressed in these treatments is the question what it means for a gravitational field
to be ``quantized''. In particular, when promoting a metric field, $g_{ab}$, to an operator, metrical properties, such 
as distances and angles, cease to have definite values until they have been ``measured'' (according to the usual Copenhagen 
view, at least), as a result of which the \emph{causal structure} of spacetime becomes indefinite.
But a crucial condition usually postulated for any quantum field theory is that of \emph{microcausality},
i.e., the property that any two quantum field variables $\Phi$, $\Phi'$ locally (anti)commute
\begin{equation}
[\Phi (x) , \Phi' (x') ]_{\pm} \; = \; 0   \qquad \quad \mbox{for} \quad (x - x')^2 > 0
\end{equation}
i.e., for points $x$, $x'$ at spacelike separation.
This local (anti)commutativity of quantum fields guarantees that local \emph{observables} constructed out of these 
fields \emph{locally commute} and this last property, together with the linearity of the standard quantum formalism,
in turn guarantees that spacelike correlations between quantum fields cannot be used for superluminal communication\footnote{See e.g. \citeN{Eberhard}.\label{nosignote}}.
So, even though it may be assumed that causality somehow becomes restored in the ``classical limit'', it is far
from clear that a theory in which $g_{ab}$ is quantized does not suffer from features usually regarded as
pathological, such as signalling.\\
On pursuing a more even-handed approach to the unification of quantum theory and general relativity, it is probably helpful that both
theories can be embedded into wider theoretical frameworks that naturally allow for amendments in each\footnote{See 
respectively \citeN{AshSch} and \citeN{Will2}.}.
What is crucially needed however are guidelines of some sort to separate physically viable amendments from non-viable ones.
This is discussed next.

\section{Empirical vs. Theoretical Principles}\label{principles}

\noindent Among the multitude of possible principles that could potentially guide the formulation of a theory of
quantum gravity, a subclass is naturally singled out by the empirical principles, i.e. those principles which are 
strongly grounded in observational facts. This division is of course not absolute and some remarks are in order.
First, there is the somewhat obvious point that even the most manifest empirical principles always exist within a 
certain theoretical context (which, according to any reasonable criterion, may still be trivial). 
Or, put slightly differently, observations, as the slogan goes, are always ``theory laden'' - something which is in 
particular true for modern precision experiments in particle physics and cosmology\footnote{For a recent discussion of this
point in the case of particle physics, see \citeN{Lyons}. Against this background, one may attempt to assign a measure of 
genuineness to empirical principles, depending on the amount of non-trivial theoretical context involved.
For instance, the equivalence principle would obviously classify as an empirical principle pur sang (by 
any reasonable measure), whereas many entries listed as observational facts in the \emph{Review of Particle Physics} would 
fall considerably short of approaching this status.
It is worth noting in this regard, that the four principles proposed in section \ref{QGprin} are also genuinely empirical
in the sense here intended.\label{moderntests}}.
For instance, a well established observational principle in cosmology is that, on the largest astronomical scales,
the universe is (spatially) homogeneous and isotropic to a very high degree\footnote{It is recalled that 
a spacetime $(M,g_{ab})$ is said to be (spatially) homogeneous if there exists a 
one-parameter family of spacelike hypersurfaces, $\Sigma_t$, $t \in \real{}$, which 
foliate the spacetime in such a way that for each $t \in \real{}$, and for any two points in $\Sigma_t$, there exists an isometry,
which maps one point into the other point. In addition, $(M, g_{ab})$ is said to
be (spatially) isotropic at each point, if there exists a timelike vector field, $u^a$
(representing the normalized tangent field of a family of observers), such that 
there exists a subgroup of the isometry group of $(M, g_{ab})$, which for each
$p \in M$ acts transitively on the space of vectors orthogonal to $u^a$ at $p$,
but which leaves $p$ and $u^a$ fixed.
In fact, spatial isotropy at each point implies spatial homogeneity, but not vice versa (cf. \citeN{Wald1}). \label{isotropynote}}. 
It is clear however, that no matter how sophisticated and accurate cosmological tests become, they will be restricted
to our own galaxy, for any foreseeable time to come. In other words, in extending the factually observed isotropy and 
homogeneity to an empirical principle for the universe at large, it is assumed that our (galaxy's) position in the 
universe is not privileged. But this assumption has of course been generally accepted as increasingly reasonable in 
science from the time of Copernicus onwards (in fact, nothing short of a scientific revolution would result were this 
assumption proven wrong one day). So, while it is certainly true that observational principles always have to be interpreted 
against a certain theoretical background, it appears slightly pedantic (or at very least scientifically counter-productive) to 
stress this too much.
A second point about empirical principles concerns the fact that if they are taken to be crucial in guiding theory selection
- as is done here for the case of quantum gravity - some readers may protest that theories are usually
\emph{underdetermined} by observational data. That is, there will always be more than one theory consistent with a given set 
of such data and an objective basis is thus lacking, so the argument goes, for choosing one particular theory from a 
number of different empirically equivalent rival theories. 
As is well known however, this argument is only effective if observational data are taken as the \emph{sole} criterion for 
theory selection - something which is certainly not intended here\footnote{Other possible criteria frequently imposed
in this regard include (amongst others) intelligibility, development potential and (mathematical) naturalness/aesthetics. For further discussion
of these issues, see e.g. \citeN{Cao}.}.
This point will be further illustrated in subsection \ref{GRandstmod} in terms of the equivalence principle and the role
it played in the development of general relativity.
A final and related issue concerns potential guiding principles which are not empirical, i.e., the ``theoretical principles''.
Such principles are supported by observations only in the most minimal sense, in that they should not lead to conflicts 
with experiments. For instance, the ``gauge principle'', which is fundamental to the standard model of particle
physics belongs to this class. As will be recalled in subsection \ref{QTandstmod}, all known, nontrivial arguments for 
this principle are purely theoretical in character (arguments, whether they be empirical or theoretical, are here referred
to as nontrivial, if they are \emph{direct} arguments; a posteriori one could always argue that some principle is
indirectly supported if it occupies a central place in some well-confirmed theory, but such arguments are regarded as trivial).
Although theoretical principles are not necessarily invalid (in fact, as just recalled, they are in general even indispensible
for empirical principles to be effective for theory construction), their a priori foundational significance should be viewed
with proper caution in any empirical science. In particular, care should be exercised to avoid what might be called
``epicycle'' principles.

\subsection{General Relativity and the Standard Model of Cosmology}\label{GRandstmod}

\noindent A prime historical example of an observational fact being paramount to the formulation of a new theory
is provided by the (weak) equivalence principle, i.e., the fact that, ignoring air resistance, (test) bodies in a 
gravitational field fall at the same rate, independent of their composition and internal structure.
As is well known, Einstein took this principle as a key hint for obtaining a description of gravitational phenomena 
in accordance with relativistic principles, but in terms of a generalized spacetime geometry 
(rather than to simply try and shoehorn these phenomena into the framework of special relativity).
It is instructive to briefly recall the essential steps involved.
Imagine oneself in the year 1905, just following the announcement of special relativity, and suppose one were to attempt
to describe gravity in terms of basic relativistic ideas.
Would it be reasonable to indeed pursue development of a \emph{special} relativistic theory of gravity ?
This may seem difficult to answer, given our present-day biasses, but the equivalence principle actually provides a 
powerful hint here.
That is, it is exactly as a consequence of this principle that an observer who is located
within an effective vacuum and who is studying the motions of test bodies inside a freely falling laboratory (treated as another 
test body, thus ignoring possible tidal effects) in the absence of non-gravitational, external forces, would validly 
classify these motions as \emph{inertial} (i.e., unaccelerated) motions.
Thus, within any pure special relativistic theory of gravity, the equivalence principle implies that there are \emph{two}
distinct classes of preferred motions in spacetime, the usual inertial motions and the ``free-falling'' motions.
Of course, the inertial motions of Minkowski spacetime precisely correspond to the geodesics of that spacetime, whereas 
the free-fall trajectories lack such a clear geometrical interpretation within a special relativistic context. 
But this suggests that spacetime is perhaps only ``locally Minkowskian'' and that gravitational effects are somehow to be 
attributed to spacetime structure, with the free-fall trajectories corresponding to the geodesics of a more general spacetime geometry. 
In fact, it is possible to considerably strengthen this intuitive argument and present a powerful case that gravity is 
necessarily a curved spacetime phenomenon if the so-called \emph{Einstein Equivalence Principle} (EEP) is satisfied\footnote{See in 
particular \citeANP{Will1} \citeyear{Will1,Will2}.
EEP consists of the (weak) equivalence principle together with two additional empirical
conditions, usually called ``Local Lorentz Invariance'' (LLI) and ``Local Position Invariance'' (LPI), and which essentially
boil down to a local restatement of the relativity postulate for (test body) systems undergoing free fall. See also \citeN{Holman2}.
Yet another equivalence principle, the so-called \emph{Strong Equivalence Principle} (SEP), consists of the extension of EEP
to include also situations involving gravitational self-interaction effects. There is some evidence that general relativity is singled
out uniquely by SEP. That is, of all so-called \emph{metric theories of gravity} - i.e., theories able to incorporate EEP -
it appears that only general relativity can also incorporate SEP.}.
Since there is overwhelming experimental evidence that the latter is indeed the case, it is thus possible to draw some
far-reaching conclusions about the physical nature of gravity, space and time on an almost purely empirical
basis as a result\footnote{For a relatively recent discussion of the experimental status of LLI and LPI, see \citeN{Will2}.
For other lines of argument casting strong doubt on the possibility of incorporating gravitation within 
the framework of special relativity, see e.g. \citeN{Penrose8}, \shortciteN{MTW}.}.\\
As is well known of course, in concretely arriving at his field equation, Einstein also made use of other, non-empirical
arguments, most notably the principle of ``general covariance'' and ``Mach's principle''.
According to the most common understanding, the principle of general covariance expresses the form invariance of the
equations of physics with respect to general coordinate transformations.
But as was already clearly realized as early as 1917, \emph{any} physical theory formulated in terms of tensor fields
is generally covariant in this sense, and so, as a criterion for theory selection, the principle of general covariance
in this particular form is rather useless.
General covariance may also be understood in a more restricted sense however, namely, as imposing the condition of 
``background independence'' (or, equivalently, ``no prior geometry'') and in this form it can (and indeed did in the
case of general relativity) act as a sieve for separating viable from non-viable theories of gravity\footnote{For some further 
discussion of these two distinct usages of the principle, see \citeN{Holman3}.}.
Similarly, Mach's principle is based on the intuitive idea that the concept of inertia should somehow \emph{emerge} 
from the various interrelationships between all material bodies present and should thus (in principle) only be meaningful 
\emph{relative} to the entire matter content of the universe (or, in terms of Wheeler's aphorism : ``mass there governs inertia here'').
Unfortunately however, it has proven difficult to implement this idea in a mathematically precise way, something 
which has generated a vast literature on the meaning and correct implementation of Mach's principle, as well as the question 
of whether general relativity is actually a Machian theory.
Nevertheless, it appears to be generally accepted nowadays that an appropriate form of Mach's principle is in fact contained within general 
relativity, as exemplified for instance by the experimentally confirmed Lense-Thirring effect (i.e. the ``dragging of inertial frames'') 
produced by rotating matter and, more generally, by the fact that spacetime geometry - as encapsulated in the metric - is 
partly determined by the matter content of spacetime via Einstein's equation.\\
The point of the foregoing remarks is to emphasize that, in addition to paying due attention to a key empirical
principle, (i.e., the equivalence principle), the choice to consider some descriptions of physical reality to be 
more intelligible than others merely on theoretical grounds (for instance, descriptions with a certain ``Machian symmetry'' 
between causal agents, as opposed to descriptions that lack such symmetry), was also of crucial importance in fleshing
out the detailed theoretical structure of general relativity.
However, although of crucial importance, the theoretical arguments were most definitely not \emph{as} important as the
empirical principle. First, this is just a \emph{historical} fact about the actual development of general relativity.
As is well known, after formulating his version of the equivalence principle (the ``happiest thought'' of his life),
Einstein went on to consider its ramifications in connection to spacetime geometry.
In particular, according to his famous rotating disc argument\footnote{\citeN{Einstein3}.}, a rigid disc at rest with
respect to a uniformly rotating coordinate system exhibits non-Euclidean features (i.e., because of Lorentz contraction, 
the ratio of the circumference of the disc to its diameter appears smaller than $\pi$ to a non-rotating observer) and the
equivalence principle then suggests that spacetime itself should have a non-Euclidean character in the presence of a 
gravitational field\footnote{This latter conclusion was reached shortly afterwards by Einstein and Grossmann \citeyear{EinGro}, 
who also generalized the flat metric, $\eta_{ab}$, of special relativity, to a general curved metric, $g_{ab}$. 
In fact, the primary role of the equivalence principle in arriving at these conclusions was explicitly acknowledged by these authors :
\begin{quote}
The theory described here originates from the conviction that the proportionality between the inertial and the gravitational
mass of a body is an exact law of nature that must be expressed as a foundation principle of theoretical physics. 
\end{quote}
(as quoted in \shortciteN{MTW}). Strictly speaking, this line of argument requires (non-empirically) extending the 
equivalence between uniformly accelerating reference systems in the absence of gravity and stationary reference systems 
located inside a uniform gravitational field to the case of arbitrary reference systems and gravitational fields. Nevertheless, it is obvious that the starting point of the whole deduction
was firmly grounded in an empirical fact, while it was seen earlier that from a modern vantage point the role of
non-empirical considerations can be substantially reduced, in as much as EEP - which is fully empirical - implies that gravitation is necessarily described by a metric theory.}.
Apart from the factual historical course of events, a different (but related) reason for attributing priority to the
equivalence principle is that it is difficult to see how to even make sense of the other, i.e. theoretical, arguments
without first adducing it as support for curved spacetime geometry (i.e., what could be the meaning of general covariance
in the sense of ``no prior geometry'' in this context ?)
Finally, it is a simple truth that, on general grounds, genuine empirical arguments (cf. footnote \ref{moderntests})
are more robust than their theoretical counterparts. That is, even though an empirical principle may, strictly speaking, break down at some point -
as experiments become more refined and accurate - it is more likely to always remain true within its proper context.
By contrast, non-empirical arguments, without denying their fruitfulness for theory development (as just pointed out
and as illustrated by the history of science more generally), are arguably more susceptible to \emph{fundamental} flaws. 
That is, although they may also be viewed as always remaining ``true'' as far as the
empirical structures entailed by them approximately remain true (i.e., in the sense just noted), they may become
deeply flawed from a foundational perspective and due caution is therefore warranted in connection to them (as 
the history of science also shows; one recalls Kant's famous ``synthetic a priori'' statements about e.g. the Euclidean nature
of space in particular).\\
A second example of an important observational principle connected with general relativity (in particular
cosmology), is the already mentioned isotropy of the universe at the largest astronomical scales.
The strongest support for this principle comes from measurements of temperature fluctuations in the 2.7 K 
\emph{Cosmic Microwave Background (CMB)}, which, according to Big Bang nucleosynthesis, is the
distribution of photons left over from ``decoupling time'', $\tau \simeq 380.000$ years, sufficiently cooled down
to compensate for the universe's overall expansion.
Indeed, the extraordinary uniformity of the CMB (i.e., to a few parts in $10^5$) implies that the universe must have
been highly isotropic at that time, as the minute temperature variations in the CMB are representative measures
to a similar degree of accuracy of matter density fluctuations at that epoch\footnote{See \shortciteN{Spergel}. 
In fact, the CMB data give rise to the most accurate agreement between an observed intensity spectrum and a theoretical 
black-body radiation curve (as given by Planck's famous formula) known in observational science.}.
The crucial point is now that isotropy is a mathematically highly restrictive condition, which fixes the cosmological 
model to be of \emph{Friedmann-Lema\^{i}tre-Robertson-Walker (FLRW) type}. The latter constitutes the so-called standard 
model of cosmology and is completely specified in terms of two fundamental equations, involving essentially only four
experimental input parameters\footnote{More precisely, for a FLRW cosmological model, Einstein's equation reduces 
to two independent components, which can be written as 
\begin{eqnarray}
1 \: + \: \frac{k}{H^2 a^2} & = & \Omega_m \: + \: \Omega_{\Lambda}                			\label{FLRWsol1}  \\
q 			    & = & \frac{1}{2} \Omega_m \: - \: \Omega_{\Lambda} \: + \: 4 \pi P/H^2 	\label{FLRWsol2}
\end{eqnarray}
where geometrized units were used (i.e., $G_{\mbox{\scriptsize N}} = c = 1$) and 
where $a := a(\tau)$ denotes an overall scale-factor of the homogeneous three-geometries, $\Sigma_{\tau}$ (cf. footnote \ref{isotropynote}), 
$H := \dot{a} / a$ denotes Hubble's ``constant'' (the dot representing a time derivative), $q := - a \ddot{a} / \dot{a}^2$ is the so-called deceleration
parameter, $k$ is a discrete parameter which equals either $-1$, $0$ or $+1$ (according to whether the surfaces of 
homogeneity are respectively hyperbolic, flat or spherical), $\Omega_m := 8   \pi   \rho / 3H^2$ and 
$\Omega_{\Lambda} := \Lambda / 3H^2$ are dimensionless parameters that respectively measure the density of matter, $\rho := \rho (\tau)$,
and the ``energy of the vacuum'', i.e., the cosmological constant, $\Lambda$, in terms of the critical density,
$\rho_c := 3H^2 / 8 \pi$, and $P$, finally, denotes the pressure of the matter distribution (which effectively vanishes 
for the present ``dust'' dominated epoch).
Many contemporary authors take the standard model of cosmology to consist of the FLRW context, together with the ideas that
(a) the very early universe went through an extremely rapid period of inflation and (b) a substantial contribution
to the overall matter and energy content of the observable universe comes from ``(cold) dark matter'' and ``dark
energy''. Given the controversial status of both inflation (as will become clear shortly) and dark matter (in the light
of the partial successes of MOND-type models), I prefer to reserve the term ``standard model'' to the FLRW context only
and to refer to the model just described as the ``concordance model'' (the latter terminology being in agreement also with common parlance).}.
Thus, given general relativity, out of all conceivable cosmological models, empirical considerations essentially
single out uniquely a particular constructive theory as the preferred model.
There is, however, a major difficulty with this model, if it is assumed that the universe was ``born'' in a completely generic state.
That is, at first glance, the extreme uniformity of the CMB appears highly suspect
because of the presence of \emph{particle horizons} (which exist ``throughout history'' in all physically relevant FLRW models).
This means in particular that there were many regions in the early universe which were out of each other's causal influence, 
although somehow the photon distributions in all these regions nevertheless ended up at the same temperature to a few parts in $10^5$ !
Now, the majority of present-day cosmologists seem to regard ``inflation'' as a satisfactory solution to this conceptual
problem, but it will be argued in subsection \ref{CMBuniformity} that inflationary models generate substantial
conceptual problems of their own and that, moreover, the successful predictions of such models do not necessarily
confirm inflation as such\footnote{Often, the standard cosmological model is said to exhibit a second major
difficulty, for which inflation is then also regularly proposed as a solution.
This is the remarkable circumstance that, on average, the universe appears very nearly \emph{(spatially) flat}.
As I have argued elsewhere however (\citeN{Holman4}), it is rather debatable (at the very least) whether this 
observational fact constitutes a genuine conceptual problem for FLRW cosmologies. See also \citeN{Rindler}.}.
In fact, it will be argued that the observed near isotropy of the large-scale universe should act as a selectional
guideline for quantum gravity and will be accounted for in terms of the latter through a lawlike initial condition.
Thus, just as the equivalence principle pointed to a conceptual anomaly within special relativistic gravitation theory,
the extreme uniformity of the CMB appears to tell us something similarly profound about fundamental physics
(according to this line of thought, inflationary models would then be quite analogous to initial attempts to shoehorn 
gravitational physics into the framework of special relativity).

\subsection{Quantum Theory and the Standard Model of Particle Physics}\label{QTandstmod}

\noindent Just as the equivalence principle does not accord well with special relativistic physics, there were
several observational facts in the late 19th, early 20th century - such as the spectral distribution laws of black-body
radiation and the interference patterns built up from weak, single-photon sources - that did not really fit into (or were even
in direct conflict with) the generally accepted theories of mechanics and electrodynamics at the time.
And just as the equivalence principle played a key role in the development of a revolutionary new theory of gravity,
spacetime and dynamics for the ``macroscopic'' physical realm, these empirical facts were paramount to the development
of a radical new theory of matter and dynamics for the ``microscopic'' physical domain, viz. quantum theory,
which finally rose on the scene in the late 1920s, as an amalgam of many earlier obtained, partial results by different 
investigators. As recollected in the introduction - and in sharp contrast to the situation for general relativity - 
however, the final form and physical interpretation of quantum theory was strongly influenced by positivist ideas, 
rather than by a desire to obtain an intelligible understanding of the quantum world at a deeper level.
Indeed, the very notion of such a world was - and still often is - generally regarded as meaningless and any attempts 
to obtain a further understanding of it were explicitly advised against.
In addition and somewhat complementary to this, in the case of quantum theory there simply was no guiding empirical
principle analogous to the equivalence principle.
For instance, Planck's quantum hypothesis, even though on extreme secure experimental footing in all relevant cases, was at
the time a complete shot in the dark and is moreover not a genuine universal principle (as observables may still have
continuous eigenvalues).
Similarly, Heisenberg's uncertainty principle is not really an empirical principle, but rather a feature implied by
the quantum formalism that is often invoked (not infrequently in a somewhat cavalier fashion), to theoretically account 
for phenomena such as virtual quanta.
The only empirical principle that played a role in the actual process of formulating quantum theory and which could 
ultimately prove \emph{somewhat} akin to the equivalence principle (even though, according to the views presented
here, it will \emph{not} eventually do so) is the \emph{principle of linear superposition}.
Indeed, although at present still at the ``pre-E{\"o}tv{\"o}s'' stage, if confirmed at a sufficiently macroscopic level
- and sophisticated experiments are currently under way to attempt to bring light into this matter (see e.g. \shortciteN{RomeroIsartcs}) -
a strong incentive would be provided to view quantum theory as the ``whole truth about the world'' and to consequently
adopt an essentially Bohrian, ``extreme statistical interpretation'' - or rather non-interpretation - of it\footnote{Indeed,
as has been emphasized by Leggett, any attempts to further elaborate upon the theory's interpretation (be it in terms
of ``consistent histories'', ``Bohmian trajectories'', or something even more esoteric) would then seem to amount to little more than ``verbal
window dressing'' (cf. pp. 79, 232 in \citeN{Schlosshauer}).
For somewhat similar views, see also the remarks of Mermin on p. 103 of the same reference.}.\\
According to the ideas underlying the present work however, a possibility that should seriously be reckoned with is that
the outcomes of such experiments will prove to be negative at some point, so that quantum theory is simply not the whole
story.
In addition to the motivations already given (i.e., the absence of a consistent theory of quantum 
gravity at present, the lack of consensus on how to interpret quantum theory and the partial incompatibility of 
its principles with those of general relativity), a reason for suspecting this to be true comes from high-energy physics.
That is to say, quantum theory was initially developed under the assumption of atomism of some form, but has subsequently
been applied to probe the structure of matter at ever deeper levels, something which resulted in the most sophisticated,
microscopic theory of matter and interactions to date, i.e. the $\mbox{SU}(3) \times \mbox{SU}(2) \times \mbox{U}(1)$ 
standard model of particle physics.
Yet, the assumption that a theoretical formalism that was essentially extracted as an ``instrument'' from empirical data
at the relatively low energies of atomic physics still successfully applies at ever higher energy scales (or even at
whatever scale quantum gravity is supposed to make its entry), each time uncovering more sublevels in the process,
resulted in a theoretical edifice that has become somewhat unwieldy\footnote{For instance, according to the
received view it is necessary to stabilize the standard model against radiative corrections if its Higgs sector
contains relatively light scalar particles (as the results from LHC now in fact seem to have 
demonstrated) and the most popular mechanism to do this apparently is supersymmetry.
This however tends to introduce a great number of extra parameters (typically more than $100$), many of which -
given the recent LHC findings - need to be fine-tuned, thereby giving rise to the very same problem that supersymmetry
was intended to resolve in the first place.}.
As a constructive quantum theory, the standard model is strongly suggested as the simplest concrete model meeting
(i) a few basic empirical facts (such as the chiral asymmetry exhibited in particle interactions, the \emph{global} $\mbox{SU}(2)$ 
weak isospin and $\mbox{SU}(3)$ colour symmetries), which partially determine the menu of relevant fields, (ii) the so-called 
gauge principle and (iii) the requirement of (perturbative) renormalizability\footnote{See e.g. \citeN{Weinberg}.}.
Now, in contrast to the principles underlying the standard model of cosmology - which, as noted previously, are empirical and
actually \emph{fix} that model - the latter two conditions are of an entirely theoretical nature.
Moreover, unlike the theoretical principles that played a role in the development of general relativity - which,
as will be recalled from the previous discussion, are still taken to be valid (in some form) for theory development today 
- the gauge principle has a somewhat dubious status in this regard, whereas it is of course well known (as in fact also follows from earlier
remarks; cf. footnotes \ref{Landaupole}, \ref{asysafe}) that renormalizability is no longer taken to constitute a very useful
selection principle these days\footnote{A frequently given argument for fundamental gauge invariance, is that it is
an implication of including a \emph{global} symmetry - the presence of which may be dictated on essentially empirical 
grounds - in the theoretical description in a manner consistent with relativity.
There are different versions of this argument - the most advanced being due to Weyl -
but none of them is actually conclusive. See e.g. \citeN{Holman2} for further discussion of these issues.
It is of course true that features such as the necessary appearance of gauge potentials in the field equations
in the non-Abelian case, or the empirical Aharonov-Bohm effect, indicate that the whole notion of gauge is far
more subtle than could have been anticipated from an investigation of merely the classical, Abelian case.
As of yet however, these features only indicate the need to better understand this notion and they do not establish a
gauge principle in any form.
Additional support for these remarks also comes from the fact that in string theory - presumably still the most popular 
approach to go beyond the standard model of particle physics at present - gauge invariance is no longer taken to constitute
a fundamental principle (cf. \shortciteN{GSW}).}.

\section{Quantum Gravity as a Principle Theory}\label{QGprin}

\noindent If the discussion in section \ref{principles} is taken to convincingly demonstrate that appropriate 
empirical principles are required for deducing the key structural properties of quantum gravity (assuming history
to be of guidance in this respect), which principles should consequently be adopted to fulfil this purpose ?
In the following subsections, four principles are proposed which (a) are on very firm ground experimentally (in the
one case where this may be less obvious, solid arguments will of course be presented), (b) arguably connect gravity 
to the quantum in a nontrivial fashion and (c) are deemed to be of crucial importance in the search for a 
theory of quantum gravity.
Moreover, in each case, it is explained, where relevant, how the principle in question could be implemented 
mathematically in such a theory (section \ref{epi}), why it is considered to be of fundamental significance and also why contemporary 
accounts of it are insufficient.
At the risk of repetitiveness, it is stressed once more that the principles listed here are not necessarily foreseen
to function as principles of quantum gravity in the true sense of the word (although they will of course have
to be accommodated in some form in any eventual formulation of the theory).

\subsection{Quantum Nonlocality}\label{qnonloc}

\noindent Probably the most enigmatic feature of the quantum world is its nonlocality.
Indeed, assuming the (sub)microscopic world of molecules, atoms, quarks, leptons, $\cdots$, to be as real as the
``ordinary'' macroscopic world directly accessible to our senses, the violation of Bell-type inequalities - exactly
in accord with the predictions of quantum theory and as confirmed by countless experimental tests performed over the
past few decades - has forced theorists to give up on the hope of ever describing this world in a fully local manner
(there are still a few minor, most presumably soon-to-be-closed loopholes to this conclusion, but with one exception,
which is briefly addressed in the next subsection - these will not be discussed here\footnote{For a recent discussion
of some generally acknowledged loopholes and how they could be closed, see e.g. \citeN{Schlosshauer}.}).
In making this statement, it is important however to distinguish between two major notions of the term ``locality''.
According to one notion, locality is synonymous with \emph{relativistic causality}, i.e., the absence of superluminal
signalling (``no-signalling''), which is usually taken to be a fundamental tenet of relativity theories (i.e., in order
to avoid certain paradoxes already referred to earlier - more on this in subsection \ref{qindet}).
A second type of locality corresponds to the concept of supervenience (or, what boils down to the same thing, separability),
i.e., the feature that the properties of a composite system fully ``supervene'' on the properties of its subsystems.
In its usual ``minimalist'' interpretation, quantum theory, by virtue of entanglement, obviously fails to be local in
this sense. What is demonstrated by Bell-type experiments is that \emph{any} more refined interpretation
of the theory, in which subsystems always have (stochastic) properties in all relevant respects, must be nonlocal
in one of the two senses just outlined, something which is generally interpreted in terms of some sort of non-separability, 
exactly as entailed by the quantum formalism\footnote{Cf. footnote \ref{nosignote}.
More precisely, the two types of locality correspond to technical conditions known in the literature as \emph{parameter} 
and \emph{outcome independence}, respectively, and the violations of these conditions are often said to be of
a ``controllable'' and ``uncontrollable'' variety, respectively. See e.g. \citeN{Bub} and references therein.}.
Furthermore, the nonlocality demonstrated is not to the maximum extent possible, i.e. as allowed by locality in the
first sense. In fact, there is a large class of no-signalling theories which exhibit nonlocal correlations stronger
than allowed for by quantum theory, but these so-called ``super-quantum'' theories do not conform to nature\footnote{The
standard way to study such theories is through introduction of the so-called nonlocal PR box - after \citeN{PopRohr}.\label{PRbox}}.
The nonlocality displayed by Bell-type experiments is indeed in very impressive agreement with the predictions 
of quantum theory and this seems to tell us something profound.\\
Even if nonlocality is an ineliminable part of quantum theory - as the best available evidence indicates it is -
how could that have anything to do with quantum gravity ?
The link is in fact quite straightforward - although hardly ever made, to my knowledge.
Indeed, according to general relativity, gravity is directly responsible for spacetime causal structure (even in
situations where the ``force'' of gravitation is not directly relevant).
So the fact that nature exhibits nonlocal correlations, precisely in accord with quantum theory (i.e., and not with
any super-quantum theory), suggests a profound connection between gravity and the quantum.
In addition, even though not in direct conflict with relativistic causality, nonlocal quantum correlations at least
bring to the surface an existing \emph{tension} between quantum theory and relativistic ideas (this is even more
so for interpretations in which state vector reduction is taken to be a real process).
It may be that this tension is again indicative of something profound and that a theory of quantum gravity is
required to ultimately resolve it.\\
The fact that quantum nonlocality is here taken as a key principle for uncovering the structure of quantum
gravity does not contradict anything that was said in subsection \ref{QTandstmod}. 
That is, even though nonlocality is an \emph{implication} of quantum theory, rather than a foundational principle, 
this does not mean that it does not have crucial foundational \emph{significance}.
According to the above discussion it clearly does and moreover not just for quantum theory\footnote{It is worth noting 
in this regard that quantum nonlocality has been proposed as a (foundational) principle by other investigators as well.
See e.g. \citeANP{PopRohr}, \emph{ibid}.}.

\subsection{Irreducible Indeterminacy}\label{qindet}

\noindent Since ancient times, thinkers have asked whether Nature is completely governed by deterministic rules, or whether 
it in fact contains processes that are \emph{intrinsically} random.
According to the classical view -- as expressed most clearly in Laplace's conception of the universe as a giant
clockwork -- true chance events do not exist and it is a commonplace that this view was decisively overthrown by quantum
theory in the early twentieth century.
Given the lack of consensus on how to interpret quantum theory however, it is reasonable to ask - especially given
the level of technicality of typical modern discussions on quantum foundations - whether such a conclusion might
perhaps not be premature.
In this respect, it is quite remarkable that it is possible to present two rather straightforward, empirical arguments,
which are at first sight completely unrelated and which \emph{both} point to fundamental randomness.
However, as also emphasized before already, it is important to take note of the fact that observational statements
are always relative to certain - usually implicit and possibly trivial - theoretical background assumptions.
Nevertheless, it will be argued that in this case the background assumptions are so natural - in fact, arguably to at
least the same extent as the Copernican assumption underlying the observed uniformity of the CMB mentioned earlier -
that to question them not only seems contrived, but also amounts, in essence, to questioning the very validity of the 
scientific method.
Consequently, speaking of irreducible indeterminacy as an \emph{empirical} principle is argued to be fully justified.
What then are the arguments that lead to this conclusion ?\\
First, as has only been clearly realized relatively recently, given relativistic causality, violations of Bell type
inequalities can in fact be utilized to generate true random numbers\footnote{\shortciteN{Pironiocs}.}.
This is based on the fact that in any deterministic theory nonlocal correlations can in principle be used for signalling\footnote{\citeN{Valentini}.}.
Hence, if relativistic causality is to be respected, the outcomes in any Bell-type experiment cannot be predetermined,
but are in fact generated at whatever event constitutes an act of measurement.
Now, it may be surmised, as is indeed often done, that arguments for nonlocality based on Bell inequality violation
always involve an assumption somewhere that a human experimenter is free to choose some particular measurement setting.
In fact, no such assumption is necessary, as a pair of random number generators may do the job, but the point
is clear : some presumption of ``free will'' or initial randomness seems to be required for the argument to work
(this was the caveat mentioned in subsection \ref{qnonloc}).\\
Although this point is not in itself contested here, there are strong grounds to discard the possible invalidity of such
a presumption as a \emph{serious} loophole in the argument for nonlocality (although it would of course make the argument 
for indeterminism circular, logically speaking - more on this shortly).
First, even though the notion of ``free will'' is frequently used in a vague sense and has probably been a source of
debates since the dawn of humanity, it is an often overlooked fact that in one important sense, this notion is central
to the whole scientific method\footnote{See also \citeN{HawEll}, p. 189, for a clear statement of this point.}.
Suppose that human experimenters always found that a certain physical magnitude (for instance, the Hubble radius,
or the level of $\mbox{CO}_2$ gas in Earth's atmosphere) were increasing.
Normal scientific practice would then lead them to infer that the physical magnitude in question \emph{is} in fact
increasing. But in a truly deterministic universe, a giant ``conspiracy'' on the part of nature could lead human
experimenters to think that there is an increase of the magnitude, when in actual fact there is none.
Now whether anyone thinks it plausible that our universe is \emph{actually} like this is perhaps best left as a personal
matter, but human scientists surely have to \emph{behave} at least as if they do not~!
A related issue is that in standard deterministic physical theories, initial data can be specified arbitrarily.
In a ``superdeterministic'' universe in which Bell inequalities are violated however, the initial conditions
(specified at some arbitrary past, spacelike Cauchy surface) would have to be delicately constrained somehow by
the very fact that Bell inequalities are always found to be violated whenever put to the test in the future.
As this essentially introduces a teleological element into physical theory, it is quite questionable what
is actually gained this way, if the objective of the whole enterprise was to retain locality to begin with.
Finally, concerning the issue of circularity in the argument for randomness, it was recently shown that, although indeed
some initial randomness is necessary - so that, strictly speaking, the argument is indeed circular - this initial
randomness can be made \emph{arbitrarily small} (instead of generation of randomness, a more
appropriate term is therefore randomness \emph{expansion})\footnote{\citeN{ColRen}.}.\\
A second argument for irreducible randomness is not (directly) related to quantum theory and does not appear to be
very well known. It is based on the simple observation that in a strictly deterministic physical theory, all of the
``biological complexity'' - i.e., as exhibited by life in its manifold appearances - clearly present in the universe today
(or rather, in one very tiny region of the universe), would also have to be present in any prior state of the universe. 
Thus, if the universe is represented mathematically as a foliation of spacelike Cauchy surfaces, $\Sigma_{\tau}$,
complete knowledge of all physical conditions at instant $\Sigma_{\tau_0}$, characterizing the present state of the
universe in all its bewildering complexity, would in principle allow one, in accordance with Laplace, to fully
``retrodict'' the state of the universe at any (non-singular) prior instant.
But this seems highly unlikely, as it would entail that all the complexity observed in the current state would for
instance have to be ``conspirationally'' hidden in the state at decoupling time, $\tau_{\mbox{\tiny d}} \simeq 380.000$ years,
since observations of the CMB tell us that this state was essentially featureless.
Moreover, upon evolving the universe further and further back in time, this complexity would still be present (conspirationally
hidden or not), at, say, $\tau_{\mbox{\tiny EW}} \simeq 10^{-12} \, \mbox{s.}$, when according to the standard model
of particle physics, the universe entered its Higgs phase and all particles of ordinary matter acquired their masses.
But regardless of whether the Big Bang actually marked a true beginning of things in some definite sense,
$\Sigma_{\tau_{\mbox{\tiny EW}}}$, for all physical purposes may be viewed as a true initial data surface (since ``before'' 
$\tau_{\mbox{\tiny EW}}$ essentially did not exist, in view of the absence of any physical time-tracking entities -
on account of all particles being massless).
The puzzle is then how to account for an initial state with that much complexity without falling back on some form
of divine intervention.\\
Although it appears to me that the argument for the presence of an irreducible statistical element within the fabric 
of nature just presented is basically correct, its main weakness is of course its qualitative character.
For instance, it is clear that the notion of bio-complexity refers to something different from the
usual notion of complexity of a system in the sense of Kolmogorov (i.e., essentially as the shortest algorithm that 
``generates'' the system when transcribed into a sequence of bits), as systems of maximum Kolmogorov complexity are
completely random - something which is evidently not true for living organisms.
In order to capture the kind of complexity encountered in living organisms some sort of trade-off is needed between complete
randomness and complete order and it appears the appropriate notion is that of ``effective complexity'', which may roughly 
be characterized as the length of a ``concise'' description of the system involved\footnote{\citeN{GellMann}, \citeN{GellMannLloyd}.
Here, care should also be exercised to distinguish between systems with a high measure of such complexity and systems which
merely have a large amount of ``logical depth'' (such as e.g. mathematical fractals). Basically, the complexity of systems 
of the latter sort is only apparent (i.e. represents some measure of the time or number of elementary operations
needed to generate the system from an ultimately simple system).
Although the issue of how much effective complexity is present in terrestrial biochemistry appears to be unsettled, it
is clear that there is a huge amount of such complexity present in biology (i.e., a huge amount of additional information
is still needed to characterize biology, given the -- essentially simple -- laws of physics and chemistry).
The overwhelming amount of variety in life on Earth is to a very great extent dependent on historical accidents, i.e.
chance events that have occurred in the course of four and halve billion years of evolution, which cannot ultimately be \emph{derived}
from lower-level theories such as chemistry or physics.}.\\
Another issue is that the foregoing argument appears inconclusive if one refuses to view $\tau_{\mbox{\tiny EW}}$
as an initial data surface for some reason (for instance, because one does not accept the particle physics explanation
for the origin of mass). However, in that case the only way of scientifically resolving the issue of 
complex ``initial'' conditions without invoking indeterminism, would be to argue that we basically live
in an infinitely cyclic universe with eternal recurrences. Such a universe might have had some plausibility in 
Nietzsche's time, but surely physical science has moved forward since then.
So, while it is true that with both above arguments for irreducible randomness, strictly speaking one cannot avoid a 
small amount of circularity, it is necessary to look at the \emph{complete} picture.
When this is done, it is clear that both arguments complement each other (although, admittedly, the first argument
is more scientifically rigorous at the present time) and that it moreover seems fully justified to actually speak 
of them as empirical arguments for indeterminism\footnote{It is of course true that the first argument crucially
depends on quantum nonlocality and since the latter was already proposed as an empirical principle previously, by
itself this argument thus does not \emph{separately} establish indeterminacy as an empirical fact.
Clearly, this is where the second argument comes in. One advantage of proceeding this way could be that if
quantum nonlocality and randomness were to both appear in the form of actual \emph{postulates} in a theory of
quantum gravity, it may be possible to establish relativistic causality as a \emph{theorem} in a manner that does 
not rely on the linearity structure of the standard quantum formalism (cf. footnote \ref{nosignote}).}.
In the context of the present discussion, an element of true chance is clearly suggestive of a more fundamental
underlying dynamics that somehow unifies the unitary evolution postulate and the projection postulate of standard
quantum theory. It is then only natural to link such putative dynamics to a theory of quantum gravity
(this will become clearer upon discussion of the remaining two empirical principles below).

\subsection{Thermodynamic Arrow of Time}\label{TDarrow}

\noindent One of the most basic facts of physical experience - familiar to anyone who has ever tried to 
watch a movie backwards - is the apparent directionality of time.
In particular, according to the (empirical) second law of thermodynamics, the entropy of an isolated system in 
general increases, but never decreases.
Yet, when it is asked where this ``thermodynamic arrow of time'' comes from, our presently most fundamental
physical theories seem unable to provide an answer, as they are (taken to be) completely time-\emph{symmetric}\footnote{Indeed, 
the conventional viewpoint appears to be that, on account of respectively (the unproven, but - within its proper context - 
physically reasonable conjecture of) strong cosmic censorship and ``post-selection'' of ensembles, general relativity and 
quantum theory (including the projection postulate) are both fully time-symmetric theories.}.
The commonly given solution to this paradox traces the observed time-irreversible behaviour to initial conditions. 
That is, a coarse-graining of phase space in terms of volumes with the same ``macroscopic appearance'' is introduced
and it is argued that, since the volume of a particular coarse-grained state (i.e., essentially the entropy of that state
exponentiated) is a measure of the fraction of time a system spends in that state, with overwhelming probability, 
physical systems tend to evolve into ever larger coarse-graining regions, until they finally enter the largest, 
i.e. maximum entropy, region, in which they then stay forever.
So on this account, the second law of thermodynamics is more or less a straightforward consequence of the fact that, 
for some reason, the very early universe happened to be in an extremely low entropy state - as firm empirical evidence
actually indicates it was\footnote{See e.g. \citeANP{Penrose1} \citeyear{Penrose1,Penrose9}, \citeANP{Wald2} \citeyear{Wald4,Wald2}, \citeN{Feynman}.}.\\
Now, it should immediately be added that the above arguments are somewhat heuristic and some qualification therefore
seems to be in order.
First, the extreme uniformity of the CMB mentioned before means that the very early universe was 
essentially in a state of thermal equilibrium and therefore, it would seem, essentially in a state of 
\emph{maximum} entropy. 
As is unfortunately not always recognized however, identification of a thermal equilibrium state with a 
state of maximum entropy is only permissible if gravity (and similar long-range effects) can safely be neglected. 
In essence, gravitational clumping \emph{always} increases entropy, so that upon following the standard qualitative
account of structure formation and stellar evolution from, say, decoupling time onwards, the universe's entropy
has since become (vastly) greater.
In fact, this process of entropy build-up is saturated by black hole formation.
That is, for a given, sufficiently large distribution of matter, the state of maximum entropy is essentially the
state in which the entire matter distribution has collapsed into a black hole.
Thus, upon estimating the amount of matter in the observable universe expected to already have collapsed into
black holes and using the Bekenstein-Hawking formula for the entropy of a black hole, it is possible to obtain
a semi-quantitative measure of the ``degree of specialness'', i.e. ultra-low entropy, of the very early universe\footnote{Taking
the estimated total entropy due to black holes, $S_0$, in the current universe as an upper bound for the entropy of the
very early universe and taking for the maximum entropy, $S_{\mbox{\tiny max}}$, the entropy of a black hole the
mass of the observable universe, Penrose has obtained an upper bound of $e^{S_0 - S_{\mbox{\tiny max}}} \simeq  10^{10^{- 123}}$  for the fractional phase
space volume representing the initial state of the universe.}.\\
A second point is that, when applied to ordinary laboratory systems (not essentially involving gravitational degrees 
of freedom), arguments about evolution in phase space of the type sketched above assume time translation invariance 
and ergodicity and it seems quite clear that these assumptions in general fail to hold within a general relativistic context.
It also seems clear however that a theory of quantum gravity is precisely what is needed to properly address these
issues. In fact, it is probably not reasonable to expect general relativity to have a sensible thermodynamics anyway,
whereas features such as the Hawking-Unruh effect and the remarkable similarity between the laws of black hole mechanics
(which are essentially rigorous theorems in differential geometry) and the (essentially empirical) laws of ordinary
thermodynamics, nevertheless strongly hint at a deep connection between general relativity, quantum theory and 
thermodynamics, that will be part of a future theory of quantum gravity.\\
Assuming that the puzzles involving gravitational entropy will eventually indeed be resolved by such a theory, the
question that obviously remains about the origins of the thermodynamic arrow of time is : \emph{why} did the
early universe have such an extremely low entropy ?
Here, two basic types of answers can be sought.
In one type of approach it is envisaged that the universe basically started off in an as generic state as possible
and the idea is to demonstrate that dynamical evolution ``generically'' produces a highly symmetric, ultra-low
entropy state within a sufficient timeframe.
However, unless a major change in the second law of thermodynamics is contemplated, this would mean that the
initial state of the (observable) universe had even lower entropy - and was therefore even more improbable.
It therefore does not appear, a priori, that this general type of answer can be successful (in fact, the
received view among contemporary cosmologists appears to be that certain inflationary models successfully address
the issue just raised, but this view will be contested in subsection \ref{CMBuniformity}).
The radically different alternative type of answer holds that - for essentially the reason just given - it would
be ill-guided to seek a dynamical explanation for the ultra-low entropy origin of the early universe.
Rather, this latter feature should be taken at face value and the most natural inference, arguably, is that
the initial entropy, for some reason currently not understood, was \emph{constrained} to be very low.
Moreover, since quantum gravity arguably \emph{does} contain a fundamental arrow of time (so as to e.g. exclude 
``white holes'', i.e., the time-reverses of black holes)\footnote{Cf. \citeN{Penrose1}.
Although white holes do not appear to be strictly forbidden, any quantum gravity theory that includes these entities 
must come to terms with a number of serious difficulties afflicted by them. 
Also, if information is truly lost in the process of black hole evaporation, CPT and time-reversal invariance would 
necessarily have to be broken in quantum gravity. See \citeN{Wald4} for a discussion of these issues.}, it seems quite 
plausible that such a constraint will be provided in the form of a \emph{lawlike} initial condition in quantum gravity.

\subsection{Uniformity of the Cosmic Microwave Background}\label{CMBuniformity}

\noindent According to the cosmological ``concordance model'' (which is generally taken by present-day
cosmologists as the best description of the available data\footnote{See for instance \citeN{Ellis2}.}), the very early universe went through an epoch
of \emph{inflation}, during which it expanded by a googol (i.e., $10^{100}$) factor within just an extremely short
timeframe ($\simeq \, 10^{-30}$ s. seems to be a typical order).
It appears that the presence of such an epoch is commonly regarded as a solution to the horizon problem mentioned in
subsection \ref{GRandstmod}, the underlying idea being basically as sketched in the previous subsection,
namely that the universe started off in a completely ``generic'' state, but that after many e-folding times of
exponential expansion, thermal processes had managed to effectively isotropize it.
As is well known however, this inflationary explanation of the near FLRW geometry of the observable universe suffers
from a number of problems.
The most significant of these is that it is very far from clear whether inflation can actually be said to achieve 
its aims. Highly special initial conditions still appear to be needed to ensure that the universe effectively 
isotropizes at sufficiently late times \emph{assuming} inflation to take place, while it is moreover far from
clear that it is actually ``probable'' for the early universe to have undergone an inflationary phase\footnote{Cf.
\citeN{Penrose3}, \citeN{Ellis1}, \citeN{Turok}, \citeN{Wald2}, \citeN{GibTur}.
Other difficulties of inflation for instance are (a) the physical nature of the scalar field that is supposed to 
drive the exponential expansion, (b) the fact that a ``generic'' initial state in a time-symmetric theory
should in general include white holes, implying gross violations of the second law of thermodynamics and
(c) the existence of deep trans-Planckian modes in chaotic inflationary models.}.\\
Somewhat ironically, inflationary models have managed to predict quite successfully minute \emph{deviations}
from exact uniformity that are well borne out by measurements of the CMB.
In particular, inflationary models generically predict that the tiny temperature fluctuations in the CMB are
Gaussian and scale-free, which is indeed what is found in precision measurements by for instance WMAP.
However, alternative cosmological models for the very early universe (without inflation) have been found 
which lead to these same predictions\footnote{Cf. \citeN{HolWald}, \citeN{SteTur}.}.
It therefore seems highly premature - especially also in view of the aforementioned foundational problems - to conclude
that the CMB precision data have actually demonstrated it likely that the early universe went through an inflationary phase.
On the other hand, on taking the CMB uniformity at face value, it might be envisaged that the universe was somehow
``born'' in an (almost) isotropic state.
Such a scenario has been argued unlikely on the grounds that any ``reasonable measure'' on the space of all geometries
assigns zero effective probability to FLRW geometries.
However, this does not appear a very convincing argument.
For instance, in view of the highly special properties of four-dimensional manifolds (i.e., $\mathbb{R}^4$ is the only
Euclidean space that admits ``non-standard differentiable structures'' and it moreover does so infinitely many times\footnote{See e.g. \citeN{DonKro}.}),
it could equally well be reasoned that it is ``extremely improbable'' that spacetime happens to be four-dimensional
and that one should therefore attempt to seek a dynamical explanation for this common fact of experience as well.
Just as with the universe's isotropy, it seems more reasonable to take the specialness of $\mathbb{R}^4$ at face value
and to attempt to find out what it can further teach us about the nature of physical reality (see section \ref{epi}
for some additional comments on this issue).
At any rate, it would appear that a lawlike initial condition within a theory of quantum gravity, that somehow
enforces isotropy, would offer a reasonable explanation of both the thermodynamic arrow of time and the 
near-perfect uniformity of the CMB.

\section{Epilogue}\label{epi}

\noindent Do the four principles listed previously naturally point to some mathematical scheme that encompasses them ?
Before addressing this it is worth noting that the second principle (indeterminacy) appears to have a somewhat special status
in the sense that it connects \emph{directly} to the three remaining principles.
The link with quantum nonlocality was already explained. As for the two other principles (which obviously also link
directly with each other), the connection arises as follows.
The extreme ``orderliness'' of the very early universe - as implied by both general reasoning based on the second law
of thermodynamics and the extra-ordinary uniformity of the CMB - means that all (quantum) gravitational degrees of freedom
were effectively absent in that epoch (since, in general relativity, gravitational degrees of freedom are represented
by the Weyl tensor, $C_{abcd}$, this has led to the conjecture that vanishing Weyl curvature must be a basic \emph{law}
within quantum gravity\footnote{\citeN{Penrose2}}).
Somehow these degrees of freedom became ``activated'' in the course of evolution, in conjunction with a loss of ``order''
(as entailed by the second law) and with the generation of true complexity (in accordance with the arguments given
in subsection \ref{qindet}).
Also worth mentioning in this regard is the fact that state vector reduction has recently been invoked in 
an attempt to account for the generation of inhomogeneities in a perfectly smooth gas (which are envisaged to then act
as seeds for the growth of further structure under the influence of gravity on cosmological time scales\footnote{\citeN{Sudarsky}.}).\\
It is not immediately clear how fundamental time asymmetry - such as for instance implied by a lawlike initial condition enforcing low entropy
(and near-uniformity) - could be implemented mathematically in quantum gravity, since the usual differential equations of mathematical physics
are time reversal invariant (although see below)\footnote{For an interesting recent, mathematically oriented discussion on the distinction
between states and laws and how it might (partially) disappear, see \citeN{Davies}.}.
As for quantum nonlocality however, it relates very naturally to the mathematical ideas of \emph{spin networks}
and \emph{twistor theory}\footnote{For reviews, see \citeANP{Penrose4} \citeyear{Penrose4,Penrose6}.}.
In fact, the latter was very much developed with quantum nonlocality acting as a guiding principle.
One of the central features of twistor theory is that spacetime is a secondary concept, which should
somehow (nonlocally) emerge from the more primitive notion of a twistor space. Basically, ordinary spacetime events are
taken to correspond to holomorphic objects (i.e., Riemann spheres) in twistor space, the basic elements of which 
represent light rays (labelled also by a helicity parameter) in ordinary spacetime.
Furthermore and quite remarkably, certain aspects of the twistor formalism seem to relate naturally to some of
the other empirical principles listed, although it is yet too early to say whether these relations really exist
or are merely apparent.
For instance, the so-called ``nonlinear graviton'' construction (which relates to a twistorial description of certain
\emph{curved}, vacuum spacetimes) seems to hint at a nonlinear modification of the quantum formalism, unifying
(deterministic) Schr\"odinger evolution with (indeterministic) state reduction, while developments to resolve the
so-called ``googly problem'' (which is the problem of finding the right-handed counterpart of the originally left-handed 
nonlinear graviton) have been tentatively described in terms of a fundamental time-asymmetry present in the formalism.
It is finally worth noting that the whole twistor approach is very much based on the fundamental fact of experience
(in accordance with the views expressed here) that ``physical space'' is three-dimensional and that there is one 
dimension of ``time''.\enlargethispage*{2cm}
In fact, the theory derives its strength from that fact, as it does not really work for any other number of spatial
and/or temporal dimensions.

\newpage

\bibliographystyle{eigen}
\bibliography{foundqg}

\end{document}